\documentclass[aps,prb,superscriptaddress,twocolumn,showpacs]{revtex4-2}
\usepackage{amsmath}
\usepackage{amssymb}
\usepackage[pdftex]{graphicx}
\usepackage{bm}
\usepackage[pdftex]{color}
\usepackage{cases}
\usepackage{multirow}
\usepackage{braket}
\usepackage{mathtools}
\usepackage{comment}
\usepackage{mathrsfs}
\usepackage[commandnameprefix=always,final]{changes}

\def\stacksymbols #1#2#3#4{\def\theguybelow{#2}
    \def\verticalposition{\lower#3pt}
    \def\spacingwithinsymbol{\baselineskip0pt\lineskip#4pt}
    \mathrel{\mathpalette\intermediary#1}}
\def\intermediary#1#2{\verticalposition\vbox{\spacingwithinsymbol
      \everycr={}\tabskip0pt
      \halign{$\mathsurround0pt#1\hfil##\hfil$\crcr#2\crcr
               \theguybelow\crcr}}}

\usepackage[pdftex]{hyperref}

\begin{document}
\title{Origin of Robust $\mathbb{Z}_2$ Topological Phases in Stacked Hermitian Systems: Non-Hermitian Level Repulsion}

\author{Zhiyu Jiang}
\affiliation{Department of Applied Physics, Hokkaido University, Kita 13, Nishi 8, Kita-Ku, Sapporo, Hokkaido 060-8628, Japan}
\author{Masatoshi Sato}
\affiliation{Yukawa Institude for Theoretical Physics, Kyoto University, Kyoto 606-8502, Japan}
\author{Hideaki Obuse}
\affiliation{Department of Applied Physics, Hokkaido University, Kita 13, Nishi 8, Kita-Ku, Sapporo, Hokkaido 060-8628, Japan}
\affiliation{Institute of Industrial Science, The University of Tokyo, 5-1-5 Kashiwanoha, Kashiwa, Chiba 277-8574, Japan}


 \begin{abstract}
Quantum spin Hall insulators, which possess a non-trivial $\mathbb{Z}_2$ topological phase, have attracted great attention for two decades.
It is generally believed that when an even number of layers of the quantum spin Hall insulators are stacked, the $\mathbb{Z}_2$ topological phase becomes unstable due to $\mathbb{Z}_2$ nature. While the counterexamples of the instability were observed in several literates, there is no systematic understanding. 
In this work, we provide a systematic understanding that the robust $\mathbb{Z}_2$ topological phase in a Hermitian system with chiral symmetry against stacking. We clarify that the robustness generally originates from level repulsion in the corresponding non-Hermitian system derived from Hermitization. We demonstrate this by treating a class DIII superconductor in 1D with $\mathbb{Z}_2$ topology and the corresponding non-Hermitian 1D system in class AII$^\dagger$ with $\mathbb{Z}_2$ point-gap topology.
 \end{abstract}

\maketitle
\section{introduction}
\label{sec:introduction}

Quantum spin Hall (QSH) insulators have attracted significant interest and spurred extensive research into the field of topological insulators over the past two decades \cite{PhysRevLett.95.146802,PhysRevLett.95.226801,PhysRevLett.96.106802,doi:10.1126/science.1148047,PhysRevB.76.075301,hsieh2008topological,PhysRevB.79.161408,PhysRevLett.106.100403,PhysRevB.85.205102,PhysRevLett.111.136804,zhao2016unexpected,liu2019superconductivity,pedramrazi2019manipulating,shumiya2022evidence}. The concept of a QSH insulator was first proposed by Kane and Mele in 2005 \cite{PhysRevLett.95.146802,PhysRevLett.95.226801}. 
They showed that a non-trivial topological phase appears in a system with Kramers pairs due to time-reversal symmetry (TRS) by considering the coupling of two Haldane models \cite{PhysRevLett.61.2015} with opposite effective magnetic fields. 
They showed that the topological phase of such a two-dimensional (2D) system hosts helical edge states of Kramers pairs, where electrons with opposite spins propagate in opposite directions without back scattering. They also proposed that the topology of such a system can be defined by a $\mathbb{Z}_2$ topological invariant. 
Because of the Kramers pairs, TRS in the QSH system is called spinful TRS.
Soon after that, following a more realistic proposal \cite{PhysRevLett.96.106802}, the QSH effect was first experimentally observed in HgTe/CdTe quantum wells \cite{doi:10.1126/science.1148047}. 
The $\mathbb{Z}_2$ topological phase has also been extended to the three-dimensional systems with Kramers pair due to TRS and experimentally observed\cite{hsieh2008topological,xia2009observation,zhang2009topological,hsieh2009observation}. The theoretical and experimental progress have highly stimulated the study of the topological phase of matter.
Beyond the well-known 2D and 3D topological insulators, other topological phases exhibiting $\mathbb{Z}_2$ topology exist, which are classified by different symmetry classes. In one-dimensional (1D) systems, $\mathbb{Z}_2$ topology can also be observed in classes DIII \cite{PhysRevB.86.184516,PhysRevB.88.134523,PhysRevB.96.161407} and D \cite{kitaev2001unpaired,PhysRevB.84.054532,PhysRevB.88.075419}. Class DIII exhibits spinful TRS, examples of which include 1D superconductors with Majorana edge states. Class D systems possess particle-hole symmetry (PHS) but lack TRS, such as 1D superconductors in the presence of a magnetic field.

In recent years, research on non-Hermitian systems has gained significant attention due to its potential for novel physical phenomena \cite{PhysRevLett.77.570,PhysRevB.56.8651,PhysRevB.58.8384,bender1999PT,PhysRevLett.106.213901,jing2017high,PhysRevLett.124.236403,doi:10.1080/00018732.2021.1876991,PhysRevA.105.053718,PhysRevA.108.023721} and rich topological phases \cite{PhysRevB.84.205128,PhysRevX.8.031079,PhysRevB.99.081302,PhysRevB.99.245116,PhysRevX.9.041015,PhysRevLett.124.086801,kawasaki2020bulk,RevModPhys.93.015005,okuma2023non,PhysRevB.108.L121302,PhysRevB.109.235408}. In Ref. \cite{PhysRevX.9.041015}, the 10-fold AZ symmetry classification was extended to non-Hermitian systems which results in increasing the symmetry classes to 38. Moreover, the energy gaps in non-Hermitian systems can be categorized into two types, refers to line gaps and point gaps \cite{PhysRevX.8.031079,PhysRevX.9.041015}. The line gap is analogous to the band gap in Hermitian systems and can result in t
 opological properties similar to those in traditional topological insulators and superconductors, such as bulk-boundary correspondence (BBC) \cite{PhysRevB.99.081302,PhysRevB.99.245116}. The point gap, on the other hand, is specific to non-Hermitian systems and captures unique topological properties that do not have Hermitian counterparts, such as non-Hermitian skin effects (NHSE) \cite{PhysRevB.97.121401,PhysRevLett.123.170401,PhysRevLett.124.086801,PhysRevResearch.2.013280,PhysRevB.102.205118,PhysRevLett.129.070401,PhysRevB.105.245143,PhysRevLett.128.223903,zhang2022universal,PhysRevResearch.5.033173,manna2023inner,PhysRevResearch.6.013213}. $\mathbb{Z}_2$ topology can also be defined in non-Hermitian systems with point gaps. For 1D systems, $\mathbb{Z}_2$ point-gap topology appears in systems with spinful TRS$^{\dagger}$ (AII$^{\dagger}$, DIII$^{\dagger}$, etc) due to the emergence of Kramers degeneracy \cite{esaki2011edge,10.1143/PTP.127.937,PhysRevX.9.041015,ghatak2019new,PhysRevB.109.235408}. A simple model is the coupled Hatano-Nelson model with spinful TRS$^{\dagger}$, in which all the eigenstates are localized at both boundaries of the system, referring $\mathbb{Z}_2$ NHSE \cite{PhysRevLett.124.086801}.

By stacking multiple 1D strips or layering 2D sheets of topological materials, a hetero structure with unique and potentially advantageous properties can be realized, known as a stacked system. 
These systems are of significant interest for both fundamental research and practical applications, such as quantum computing \cite{cai2023signatures,zhong2023towards}, due to the novel electronic, magnetic, and optical behaviors that can arise from interactions between the layers \cite{rasche2013stacked,inoue2013electromagnetic,PhysRevB.99.165418,li2023tunable}.
It is generally believed that when an even number of layers where each layer has a non-trivial $\mathbb{Z}_2$ topological phase, such as quantum spin Hall insulators, are stacked, the $\mathbb{Z}_2$ topological phase becomes unstable due to $\mathbb{Z}_2$ nature. 
However, several researches shows that $\mathbb{Z}_2$ topology exhibits robustness against stacking in Hermitian systems in certain parameter regions\cite{PhysRevB.84.054532,PhysRevB.92.235407,PhysRevB.94.235414}. Despite these findings, the reason has not seriously yet discussed and there has been no systematic research identifying the underlying causes of this robustness.

In this work, we provide a systematic understanding that the robust $\mathbb{Z}_2$ topological phase in a Hermitian system with chiral symmetry against stacking. We clarify that the robustness generally originates from level repulsion in the corresponding non-Hermitian system derived from Hermitization\cite{Feinberg}. We demonstrate this by treating a class DIII superconductor in 1D with $\mathbb{Z}_2$ topology and the corresponding non-Hermitian 1D system in class AII$^\dagger$ with $\mathbb{Z}_2$ point-gap topology as an example. For the latter system, in the case of stacking two 1D chains, the four-fold degeneracy of the spectrum breaks down to two-fold degeneracy due to the level repulsion between two Kramers pairs as expected. Remarkably, $\mathbb{Z}_2$ point-gap topology at the energy $E$ in point gaps emerged as the level repulsion remains non-trivial. Moreover, through Hermitization, the energy $E$ of the non-Hermitian system takes the role of the chemical potential $\mu$ of 
 the Hamiltonian for the DIII superconductors. Due to this correspondence,
  the energy region for the non-trivial $\mathbb{Z}_2$ point-gap topology coincides with the range of $\mu$ where $\mathbb{Z}_2$ topological phase of DIII stacked superconductor is non-trivial and zero-energy states appear. Our result provides the systematic understanding of robust $\mathbb{Z}_2$ topology in Hermitian systems with chiral symmetry, as the level repulsion in the corresponding non-Hermitian systems. 
We note that the stacking 2D $\mathbb{Z}_2$ topological insulators were studied from the view point of the robustness of 2D surface states on a 3D weak topological insulator\cite{PhysRevLett.98.106803,PhysRevB.86.045102,PhysRevLett.110.236803,PhysRevB.76.075301}. However, the weak $\mathbb{Z}_2$ topological invariant for the 3D system was studies in these works and not the (strong) $\mathbb{Z}_2$ topological invariant for the 2D system.


This paper is organized as follows. In Sec. \ref{sec:section2}, we provide our main idea that non-Hermitian level repulsion can be considered the origin of robust $\mathbb{Z}_2$ topological phases in stacked Hermitian systems via a Hermitization process. In Sec. \ref{sec:model}, we define a 1D DIII superconductor and show that it can be transformed into a Hermitized version of a coupled Hatano-Nelson model in class AII$^{\dagger}$. We demonstrate that the emergence of zero-energy states in the Hermitian system can be predicted by relating the energy point $E$ in the non-Hermitian system to the chemical potential $\mu$ in the Hermitian system. In Sec. \ref{sec:non-Hermitian system}, we investigate the robustness of the $\mathbb{Z}_2$ point-gap topology against the stacking of coupled Hatano-Nelson models. We find that this robustness results from non-Hermitian level repulsion. Additionally, we use three-chain and four-chain stacked system as examples to demonstrate that the regions of
  $\mathbb{Z}_2$ trivial and non-trivial point-gap topology on the complex plane alternate along the real axis in multi-chain stacked systems. In Sec. \ref{sec:Hermitian system}, we examine the stacked DIII superconductor and confirm its robustness using the perspective introduced in Sec. \ref{sec:model}.

\section{Origin of robust $\mathbb{Z}_2$ topology in stacked systems}
\label{sec:section2}

First, we explain our main idea why the robustness of $\mathbb{Z}_2$ topological phases in Hermitian systems can be related to non-Hermitian systems.
To establish connections between Hermitian and non-Hermitian systems, the Hermitization procedure of a non-Hermitian Hamiltonian is crucial.  For any non-Hermitian $\mathcal{H}$, it can be transformed into a Hermitian Hamiltonian $H$ in the doubled Hilbert space through Hermitization\cite{Feinberg} as follows:
\begin{equation}
H=\begin{pmatrix} 0 & \mathcal{H}-E_{p} \\ \mathcal{H}^{\dagger}-E_{p}^{*} & 0 \end{pmatrix},
\label{eq:Hermitization}
\end{equation}
where $E_{p}$ represents a reference point on the complex energy plane of $\mathcal{H}$, while it becomes a system parameter for $H$.
This relation inevitably introduces chiral symmetry for the Hermitian Hamiltonian $H$, characterized by $\Gamma H \Gamma^{-1}=-H$ with $\Gamma=\tau_z$, where the Pauli matrices $\tau_{x,y,z}$ act on the doubled space. Under periodic boundary conditions (PBC), the presence of a $\mathbb{Z}$ (or $\mathbb{Z}_2$) point gap in the non-Hermitian Hamiltonian $\mathcal{H}$ implies that a corresponding real energy gap will open in the Hermitian Hamiltonian $H$, which induces edge states, and vice versa \cite{PhysRevX.8.031079,PhysRevX.9.041015,PhysRevLett.124.086801}. 

We begin with a non-Hermitian Hamiltonian $\mathcal{H}(k)$ in class AII$^\dagger$ that exhibits non-trivial $\mathbb{Z}_2$ point-gap topology.
The Hamiltonian $\mathcal{H}$ exhibits spinful TRS$^{\dagger}$, given by
\begin{equation}
\mathcal{T}\mathcal{H}(k)^{T}\mathcal{T}^{-1}=\mathcal{H}(-k),\quad \mathcal{T}=\sigma_y,
\label{eq:TRS^dagger}
\end{equation}
where the Pauli matrices $\sigma_{x,y,z}$ and a $2\times2$ matrix $\sigma_0=\text{diag}\{1,1\}$ act on the spin space.
The condition $\mathcal{T}\mathcal{T}^*=-1$ leads to the important observation that for any eigenenergy $E$, there must be at least double degeneracy, with the eigenstates $|\psi\rangle$ and $\mathcal{T}^{T}|\psi\rangle^*$, referred to as Kramers pairs. if a non-Hermitian Hamiltonian only satisfy Eq.\ (\ref{eq:TRS^dagger}), the system belongs to class AII$^\dagger$. The topological invariant of such a system can be defined by a $\mathbb{Z}_2$ invariant $\nu\in\{ 0,1 \}$, while $\mathcal{H}(k)$ exhibits non-trivial $\mathbb{Z}_2$ point-gap topology at the reference point $E_p$ when $\nu(E_p)=1$.

\begin{figure}
\centering
\includegraphics[width=\columnwidth]{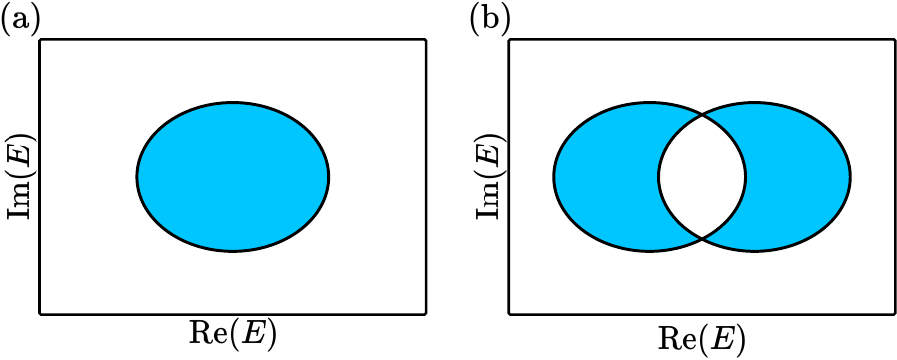}
\caption{Schematic views of the energy spectra of a stacked non-Hermitian Hamiltonian $\mathcal{H}_{\text{s}}(k)$ (a) without and (b) with coupling terms. The black solid curve represents the four-fold and two fold degenerated energy spectrum in (a) and (b) respectively. The blue regions indicate emergence of non-trivial $\mathbb{Z}_2$ point-gap topology. 
} 
\label{fig:spectrum_repulsion}
\end{figure}

For the stacked system of $\mathcal{H}(k)$, written as
\begin{equation}
\mathcal{H}_{\text{s}}(k)=\begin{pmatrix} \mathcal{H}(k) & \delta_0 \sigma_0 + i \boldsymbol{\delta}\cdot \boldsymbol{\sigma} \\ \delta_0 \sigma_0 - i \boldsymbol{\delta}\cdot \boldsymbol{\sigma} & \mathcal{H}(k) \end{pmatrix},
\label{eq:stacked Hamiltonian}
\end{equation}
where $\delta_0$ and $\boldsymbol{\delta}=(\delta_x,\delta_y,\delta_z)$ denotes the $k$-independent stacking strengths. The stacked Hamiltonian preserves spinful TRS$^\dagger$ with the symmetry operator $\mathcal{T}=\eta_0\otimes \sigma_y$, where the $2\times2$ matrix $\eta_0=\text{diag}\{1,1\}$ acts on the stacked space. One might naturally assume that $\mathcal{H}_{\text{s}}$ should be $\mathbb{Z}_2$ trivial even if each $\mathcal{H}(k)$ is $\mathbb{Z}_2$ non-trivial because of the $\mathbb{Z}_2$ nature. However, the stacking term introduces the level repulsion which makes the four-fold degeneracy of the spectrum the double degeneracy, as illustrated in Fig. \ref{fig:spectrum_repulsion}. 
In case of Fig. \ref{fig:spectrum_repulsion} (a) without couplings, since the system consists of two independent layers with non-trivial $\mathbb{Z}_2$ invariant, $\mathbb{Z}_2$ invariant remains non-trivial. In case of Fig. \ref{fig:spectrum_repulsion} (b) with finite couplings, the four-fold degeneracy is down to two-fold degeneracy due to level repulsion. Since the energy spectra winds around a point in the middle White region twice, $\mathbb{Z}_2$ invariant is trivial. On the other hand, since the energy spectra winds around a point in the blue regions once, $\mathbb{Z}_2$ invariant is non-trivial. 
Moreover, the separation distance of the energy spectrum can increase or decrease depending on the stacking strength, leading to the expansion or contraction of the $\mathbb{Z}_2$ non-trivial region.

By employing the Hermitization transform, we can extend conclusions from non-Hermitian systems to Hermitian systems. Since the $\mathbb{Z}_2$ point-gap topology in a non-Hermitian system is robust against stacking due to level repulsion, the corresponding Hermitian Hamiltonian, with parameter $E_p$ (the complex energy in the $\mathbb{Z}_2$ non-trivial region of the non-Hermitian system), also exhibits such robustness. In the Hermitian case, since $E_p$ is a parameter, we cannot understand the robustness through level repulsion as we do in the non-Hermitian case. This approach offers a clear and visual way to understand the robustness of $\mathbb{Z}_2$ topology from a non-Hermitian perspective.

\section{Demonstration: 1D DIII Hermitian Hamiltonian and 1D AII$^\dagger$ non-Hermitian Hamiltonian}
\label{sec:Hermitization}

In this section, we demonstrate our idea that the robustness of $\mathbb{Z}_2$ topological phases in Hermitian systems originates from the level repulsion of non-Hermitian systems.
To this end, we focus on a 1D topological superconductor in class DIII which possesses chiral symmetry resulting in $\mathbb{Z}_2$ topological phases and Majorana Kramers pairs. We demonstrate that the Majorana zero-energy modes are robust in such a system due to the non-Hermitian level repulsion, as discussed in Sec. \ref{sec:section2}.

\subsection{Models}
\label{sec:model}

To begin with, we consider a 1D topological superconductor in class DIII which possesses $\mathbb{Z}_2$ topological phases. Then, we derive the non-Hermitian Hamiltonian through the Hermitization in Eq. (\ref{eq:Hermitization}). 

The Hamiltonian in class DIII is formulated using fermionic annihilation and creation operators $c_{i,\sigma}$ and $c_{i,\sigma}^{\dagger}$, respectively, at a site $i$ with spin $\sigma=\uparrow,\downarrow$ and can be expressed as follows:
\begin{equation}
\begin{split}
H=&\sum_{i\in\mathbb{Z}}\left\{\sum_{\sigma=\uparrow,\downarrow}t\left(c_{i+1,\sigma}^{\dagger}c_{i,\sigma}+\text{H.c.}\right)-\mu c_{i,\sigma}^{\dagger}c_{i,\sigma} \right.\\
&+\frac{1}{2}\left[\alpha_R\left( c_{i+1,\downarrow}^{\dagger}c_{i,\uparrow}-c_{i,\downarrow}^{\dagger} c_{i+1,\uparrow} \right)\right.\\
&-d_x\sum_{\sigma=\uparrow,\downarrow}\left( c_{i,\sigma}^{\dagger}c_{i+1,\sigma}^{\dagger}-c_{i+1,\sigma}^{\dagger}c_{i,\sigma}^{\dagger} \right)\\
&+\left. \left. d_z\left( c_{i,\uparrow}^{\dagger}c_{i+1,\downarrow}^{\dagger}-c_{i+1,\uparrow}^{\dagger}c_{i,\downarrow}^{\dagger} \right)+\text{H.c.} \right]\right\}.
\end{split}
\label{eq:ts}
\end{equation}
The parameter $t$ and $\mu$ represent the hopping amplitude and the chemical potential, respectively. Additionally, $\alpha_R$ and $d_{x/z}$ represent the strengths of the spin-orbit coupling and superconductor pairing, respectively. All these parameters are real.

To study the topological properties, we diagnolize the Hamiltonian to momentum space by Fourier transform \chreplaced{$c_{i,\sigma}=\frac{1}{\sqrt{N}} \sum_k e^{ikx_i}c_{k,\sigma}$}{$c_{i,\sigma}=\frac{1}{\sqrt{N}}e^{ikx_i}c_{k,\sigma}$}, then we obtain a BdG Hamiltonian written as
\begin{equation}
\begin{split}
H_{\text{BdG}}(k)&=\Psi^{\dagger} H(k) \Psi \\
&=\Psi^{\dagger} \begin{pmatrix} h(k) -\mu \sigma_0 & i\boldsymbol{d}(k) \cdot \boldsymbol{\sigma} \sigma_y  \\ -i\boldsymbol{d}^*(k) \cdot \boldsymbol{\sigma} \sigma_y & -h^T(-k) + \mu \sigma_0 \end{pmatrix} \Psi,
\end{split}
\end{equation}
where
\begin{equation}
\Psi=\left( c_{k,\uparrow},c_{k,\downarrow},c_{-k,\uparrow}^{\dagger},c_{-k,\downarrow}^{\dagger} \right)^{T}, 
\end{equation}
\begin{equation}
h(k)=2t \cos (k) \sigma_0+\alpha_R \sin (k) \sigma_y, 
\label{eq:h(k)}
\end{equation}
\begin{equation}
\boldsymbol{d}(k)=i\left( d_x,0,d_z \right) \sin k.
\label{eq:d(k)}
\end{equation}
Here, $\boldsymbol{\sigma}=(\sigma_x,\sigma_y,\sigma_z)$ and $\sigma_0$ acting on the spin space.
The effective Hamiltonian $H(k)$ can be rewritten as
\begin{equation}
\begin{split}
H(k)=&\left(2t \cos k-\mu\right)\tau_z \otimes \sigma_0+\alpha_R \sin (k) \tau_z \otimes \sigma_y \\
&+d_x \sin (k) \tau_y \otimes \sigma_z-d_z \sin (k) \tau_y \otimes \sigma_x,
\end{split}
\label{eq:rewritten Hamiltonian}
\end{equation}
where the Pauli matrices $\tau_{x,y,z}$ and $\tau_0$ act on the particle-hole space.
The Hamiltonian satisfies the following symmetry relations
\begin{align}
\text{TRS}:\quad &TH^*(k)T^{-1}=H(-k),  \quad T=\tau_z\otimes \sigma_y, \\
\text{PHS}:\quad &PH^*(k)P^{-1}=-H(-k),  \quad P=\tau_y\otimes \sigma_y,\\
\text{CS}:\quad &\Gamma H(k)\Gamma^{-1}=-H(k), \quad \Gamma=\tau_x\otimes \sigma_0,
\end{align}
which results in that the system belongs to DIII symmetry class.

Now we derive a non-Hermitian Hamiltonian from the above DIII Hermitian Hamiltonian through a {\it inverse-Hermitization procedure}.
We consider a unitary transformation after which the chiral symmetry operator exhibits a diagonal form $\Gamma=\tau_z\otimes \sigma_0$. The transformed Hamiltonian is given by
\begin{equation}
\begin{split}
H'(k)&=e^{-i\frac{\pi}{4} \tau_y\otimes\sigma_0} H(k) e^{i\frac{\pi}{4} \tau_y\otimes\sigma_0} \\
&=\begin{pmatrix}
 & \mathcal{H}(k)-\mu \\
 \mathcal{H}^{\dagger}(k)-\mu & 
\end{pmatrix},
\end{split}
\end{equation}
where the off-diagonal matrix
\begin{equation}
\mathcal{H}(k)=\begin{pmatrix} 2t\cos k-id_x\sin k & -i(\alpha_R-d_z)\sin k \\ i(\alpha_R+d_z)\sin{k} & 2t\cos k+id_x\sin k \end{pmatrix},
\label{eq:mathbbH(k)}
\end{equation}
can be regarded as a non-Hermitian Hamiltonian $\mathcal{H}(k)$ of coupled Hatano-Nelson model \cite{PhysRevLett.124.086801}. 
The diagonal elements of $\mathcal{H}(k)$ represent the Hamiltonian of a Hatano-Nelson model without disorder and its time-reversal partner:
\begin{gather}
\mathcal{H}_{\text{HN}}(k)=\left( t-\frac{d_x}{2} \right)e^{ik}+\left( t+\frac{d_x}{2} \right)e^{-ik},\\
\mathcal{H}_{\text{HN}}^T(-k)=\left( t-\frac{d_x}{2} \right)e^{-ik}+\left( t+\frac{d_x}{2} \right)e^{ik},
\end{gather}
where the prefactors $t \pm d_x/2$ represent the asymmetric hopping terms, and the off-diagonal elements denote the symmetry-protected couplings.
We observe that $\mathcal{H}(k)$ exhibits spinful $\text{TRS}^{\dagger}$, following Eq. (\ref{eq:TRS^dagger}), with $\mathcal{T}=\sigma_y$. As a result, the energy spectrum is always double degenerate. Such a system belongs to $\text{AII}^{\dagger}$ symmetry class with $\mathbb{Z}_2$ point-gap topology according to the symmetry classification in non-Hermitian systems\cite{PhysRevX.9.041015}. The $\mathbb{Z}_2$ topological invariant $\nu$ at reference point $E_p$ on $\mathcal{H}(k)$ is defined by
\begin{equation}
\begin{aligned}
(-1)^{\nu(E_p)} & =\operatorname{sgn}\left\{\frac{\operatorname{Pf}\left[\left(\mathcal{H}(\pi)-E_p\right) \mathcal{T}\right]}{\operatorname{Pf}\left[\left(\mathcal{H}(0)-E_p\right) \mathcal{T}\right]}\right. \\
& \left.\times \exp \left[-\frac{1}{2} \int_{k=0}^{k=\pi} d k \frac{\partial \ln \operatorname{det}\left[\left(\mathcal{H}(k)-E_p\right) \mathcal{T}\right]}{\partial k}\right]\right\}.
\end{aligned}
\label{eq:Z2 invariant}
\end{equation}

\begin{figure}
\centering
\includegraphics[width=\columnwidth]{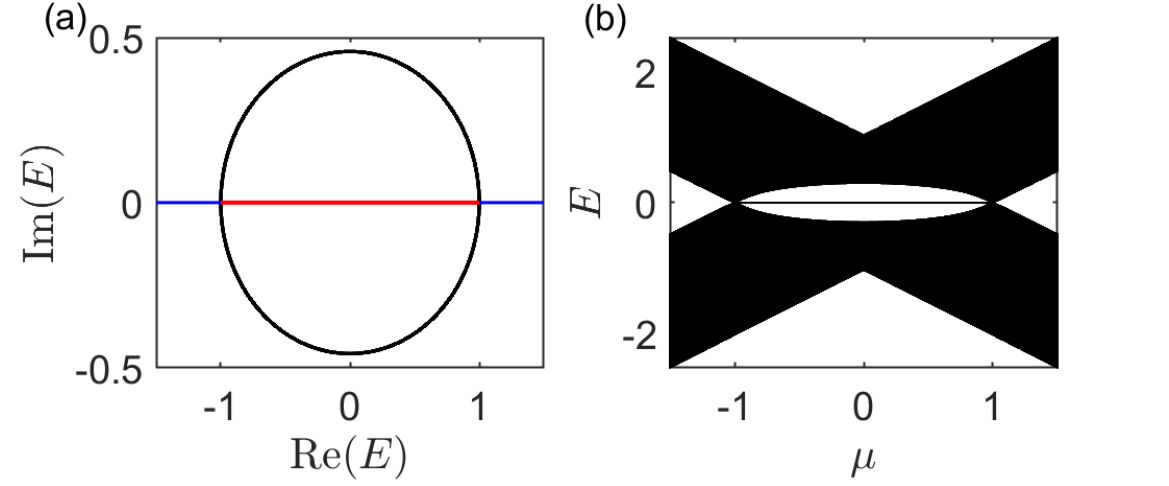}
\caption{Energy spectrum of (a) non-Hermitian coupled HN model $\mathcal{H}(k)$ and (b) corresponding DIII superconductor $H(k)$, with $t=0.5$, $\alpha_R=0.2$, $d_x=0.4$ and $d_z=0.3$. (a) $\mathbb{Z}_2$ topological invariant $\nu=1$ inside the closed curve of the complex spectrum. (b) In the region of $\mu$ encircled by the complex spectrum in (a), edge states appear in the DIII superconductor. }
\label{fig:spectrum of non-stacked system}
\end{figure}

When $\alpha_R^2<d_z^2+d_x^2$, as shown in Fig.\ \ref{fig:spectrum of non-stacked system} (a), the spectrum for the system with PBC (PBC spectrum, in short) forms a closed ellipse on the complex plane keeping the double degeneracy, leading to the $\mathbb{Z}_2$ point-gaps open at the point $E_p$ enclosed by the spectrum. Under PBC, when a point-gap opens for non-Hermitian Hamiltonian $\mathcal{H}(k)$, a real energy gap opens for DIII Hermitian Hamiltonian $H(k)$ if the chemical potential $\mu (\in\mathbb{R})$ takes a value within the point gap, as shown by the red line segment in Fig. \ref{fig:spectrum of non-stacked system}(a). In this case, $H'(k)$ [also $H(k)$] has non-trivial $\mathbb{Z}_2$ topology,  leading to the emergence of Majorana zero-energy modes under open boundary conditions (OBC) according to the BBC, as shown in Fig. \ref{fig:spectrum of non-stacked system}(b).

\subsection{Robust $\mathbb{Z}_2$ Point-gap Topology: Non-Hermitian System}
\label{sec:non-Hermitian system}

Here, we study the robustness of $\mathbb{Z}_2$ point-gap topology of the non-Hermitian system against stacking of multiple 1D chains using the coupled Hatano-Nelson model defined by Eq. \ref{eq:mathbbH(k)}. According to Eq. (\ref{eq:stacked Hamiltonian}), we define the stacked system $\mathcal{H}_{\text{s}}(k)$ as
\begin{equation}
\mathcal{H}_{\text{s}}(k)=\begin{pmatrix} \mathcal{H}(k) & \delta_0 \sigma_0+i\boldsymbol{\delta} \cdot \boldsymbol{\sigma} \\ \delta_0 \sigma_0-i\boldsymbol{\delta} \cdot \boldsymbol{\sigma} & \mathcal{H}(k) \end{pmatrix},
\label{eq:stacked coupled HN}
\end{equation}
where, $\delta_0$ and $\boldsymbol{\delta}=(\delta_x,\delta_y,\delta_z)$ represent the strength of coupling of two 1D chains. The latter one involves the spin-dependent coupling.

For simplicity, we consider the case $\delta_x=\delta_y=0$ hereafter.
We note that, to keep the symmetry class with stacking effects, in general, 
the stacking term $\sigma_0$ and at least one component of $\boldsymbol{\sigma}$ should be finite. Otherwise, the Hamiltonian in Eq.\ (\ref{eq:stacked coupled HN}) can be decomposed into two independent systems and spinful TRS$^{\dagger}$ may break, as explained in Appendix \ref{app:app2}.  Under this simplification, the PBC spectrum can be solved as 
\begin{equation}
E_{\eta\epsilon}(k)=2t\cos k+i\eta r(k)e^{i\varepsilon\phi(k)},
\label{eq:eigenvalue of stacked coupled HN model}
\end{equation}
where
\begin{equation}
\begin{split}
r(k)=&\left\{ \left[ (\alpha_R^2+d_x^2-d_z^2)\sin^2 k-(\delta_0^2+\delta_z^2) \right]^2 \right. \\
& \left. +4\sin^2 k\left[ \delta_0^2(\alpha_R^2+d_x^2-d_z^2)+\delta_z^2d_x^2 \right] \right\}^{\frac{1}{4}},
\end{split}
\end{equation}
\begin{equation}
\begin{split}
\phi(k)&=\frac{1}{2}\arctan \frac{2\sin k\sqrt{\delta_0^2(\alpha_R^2+d_x^2-d_z^2)+\delta_z^2d_x^2}}{(\alpha_R^2+d_x^2-d_z^2)\sin^2 k-(\delta_0^2+\delta_z^2)}\\
&\in\left( -\frac{\pi}{2},\frac{\pi}{2} \right).
\end{split}
\end{equation}
Here, $\eta,\varepsilon=\pm$ represent the four energy bands $E_{\pm\pm}$ of $\mathcal{H}_{\text{s}}(k)$.
Rewriting Eq. (\ref{eq:eigenvalue of stacked coupled HN model}), we see that
\begin{equation}
E_{\eta\varepsilon}(k)=2t\cos k-\eta\varepsilon r(k)\sin \phi(k)+i\eta r(k)\cos \phi(k).
\end{equation}
Note that $\cos \phi(k)\ge0$.

\begin{figure}
\centering
\includegraphics[width=\columnwidth]{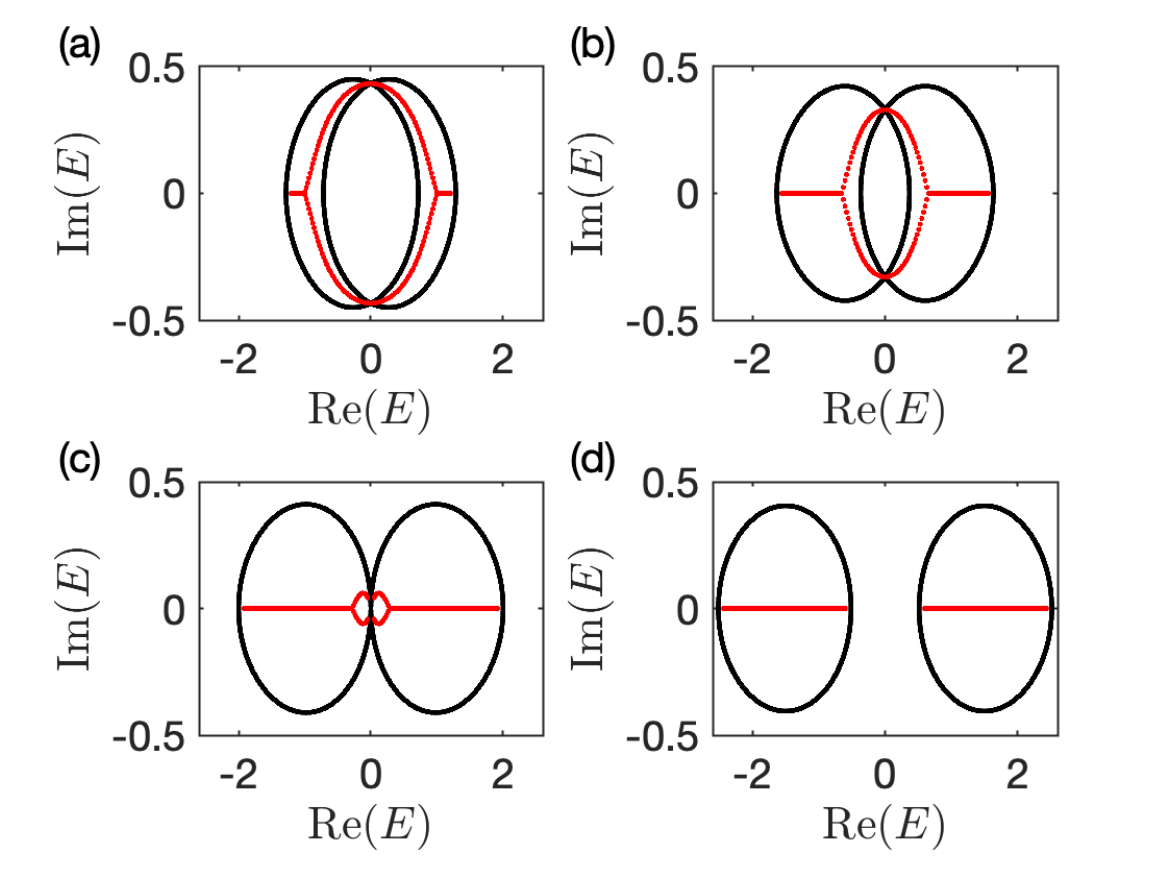}
\caption{ Complex energy spectrum of stacked coupled HN model $\mathcal{H}_{\text{s}}$ under PBC (black curves) and OBC (red curves), with parameters $t=0.5$, $\alpha_R=0.2$, $d_x=0.4$, $d_z=0.3$, $\delta_0=0.2$ and (a) $\delta_z=0.2$ (b) $\delta_z=0.6$ (c) $\delta_z=\sqrt{0.96}$ (d) $\delta_z=1.5$.}
\label{fig:spectrum of stacked coupled HN}
\end{figure}

For the stacking term $\delta_0=\delta_z=0$, there is no coupling in the stacked system, resulting in 4-fold degeneracy of the eigenvalues. In this case, the $\mathbb{Z}_2$ topological invariant $\nu$ in Eq.\ (\ref{eq:Z2 invariant}) becomes trivial due to $\mathbb{Z}_2$ nature. On the other hand, when $\delta_0\neq 0$ or $\delta_z\neq 0$, $\phi$ is finite in general. which results in lifting the four-fold degeneracy to double degeneracy because we see $E_{\eta+}=E_{\eta-}$ while $E_{+\varepsilon}\neq E_{-\varepsilon}$). This indicates that the level repulsion in $\mathcal{H}_{\text{s}}(k)$ is induced by the coupling of two chains.
In Fig.\ \ref{fig:spectrum of stacked coupled HN}, the black curves shows the PBC spectrum in Eq.\ (\ref{eq:eigenvalue of stacked coupled HN model}) for various values of the stacking strength $\delta_z$.
As schematically explained in Fig.\ \ref{fig:spectrum_repulsion}, we observe that the single closed ellipse with no stacking is split into two ellipses at the finite value of $\delta_z$. 
Especially, the regions surrounded by split spectra expand as increasing $\delta_z$. 
Remarkably, we confirmed that the $\mathbb{Z}_2$ topological invariant $\nu(E_p)$ in Eq.\ (\ref{eq:Z2 invariant}) becomes non-trivial when the reference point $E_p$ locates in the expanded regions, that is, the inside the ellipses except the common regions of inside of the two ellipses.
In particular, when $\delta_z\ge\sqrt{(2t)^2-\delta_0^2}$, all the regions enclosed by the spectrum exhibits non-trivial $\mathbb{Z}_2$ topology [Fig.\ \ref{fig:spectrum of stacked coupled HN} (c) and (d)].

Next, we verify the BBC of the $\mathbb{Z}_2$ point-gap topology in the stacked systems.
As shown by the red curves in Fig. \ref{fig:spectrum of stacked coupled HN}, which represent numerically calculated eigenvalues for the system with OBC (OBC spectra, in short), 
OBC spectra always emerge in the topologically non-trivial regions.  Therefore, the BBC of the $\mathbb{Z}_2$ point-gap topology holds in the stacked systems.

\begin{figure}
\centering
\includegraphics[width=\columnwidth]{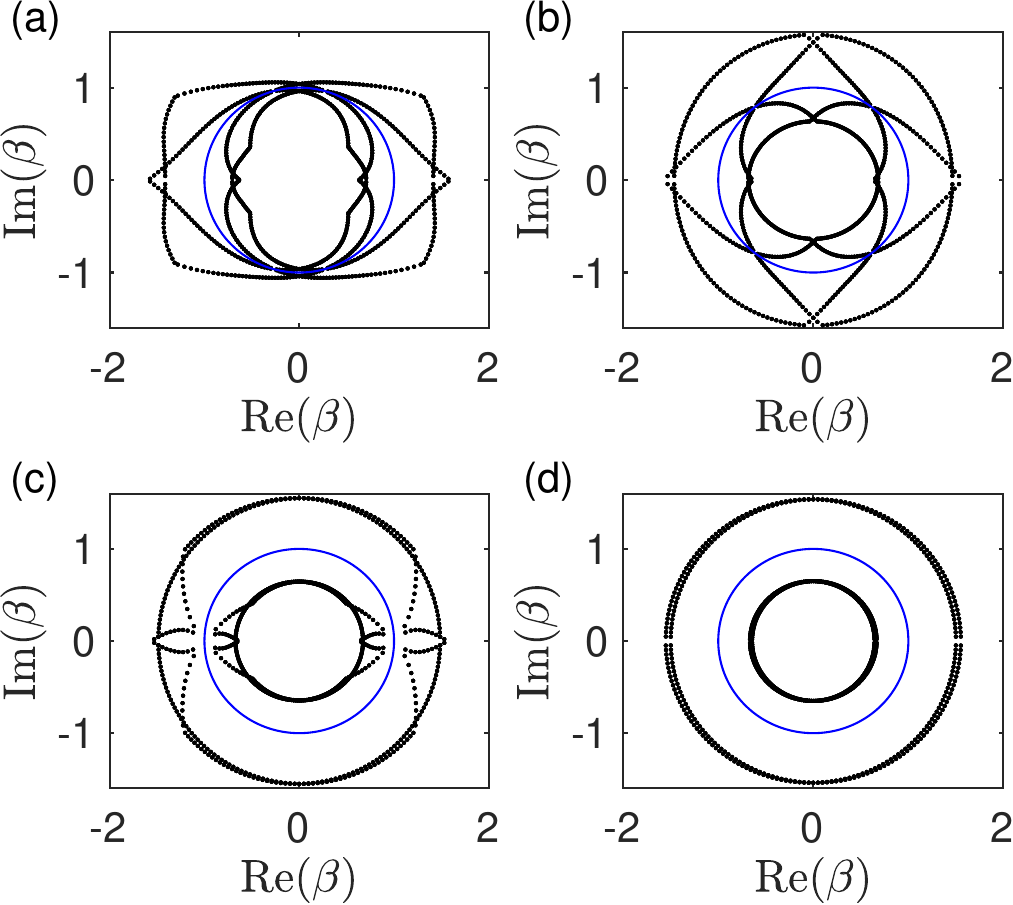}
\caption{ GBZ (black dots) and BZ (blue unit circle) of stacked coupled HN model $\mathcal{H}_{\text{s}}$, with parameters $t=0.5$, $\alpha_R=0.2$, $d_x=0.4$, $d_z=0.3$, $\delta_0=0.2$ and (a) $\delta_z=0.2$ (b) $\delta_z=0.6$ (c) $\delta_z=\sqrt{0.96}$ (d) $\delta_z=1.5$.}
\label{fig:GBZ}
\end{figure}

As a technical note, when we calculate the OBC spectra, the strong finite-size effects make it impossible to obtain the OBC spectrum by directly diagonalizing $\mathcal{H}(k)$ in real space. Therefore, we apply the generalized Brillouin zone (GBZ) \cite{PhysRevLett.121.086803,PhysRevLett.123.066404} approach according to non-Bloch band theory in the symplectic class \cite{PhysRevB.101.195147}. The GBZ corresponding to Fig. \ref{fig:spectrum of stacked coupled HN} is shown in Fig. \ref{fig:GBZ}, by the black dots, while the unit circle (blue solid curve) represents the ordinary Brillouin zone (BZ). The points on the GBZ inside of the BZ correspond to eigenmodes localized at the left boundary, while those outside of
  BZ represent right-boundary-localized eigenmodes. From Fig. \ref{fig:GBZ}(a), for a small value of $\delta_0$ and $\delta_z$, the GBZ begins to deviate from the BZ along the real axis, and the eigenmodes, especially those near the imaginary axis, are not strongly bound by the boundaries. However, for a strong stacking strength, all the eigenmodes become localized at both boundaries, as shown in Fig. \ref{fig:GBZ}, indicating the emergence of a strong $\mathbb{Z}_2$ NHSE. 
We note that the robustness of $\mathbb{Z}_2$ invariant is rather the general property of the stacked Hamiltonian belonging to class AII$^\dagger$. The same results are obtained even by using a different model as shown by Appendix \ref{sec:appendixA}.

Furthermore, the robustness of the $\mathbb{Z}_2$ point-gap topological phase due to the level repulsion can be extended to multi-chain stacked systems. It is expected that, for an $n$-chain stacked system, the energy spectrum splits from a $2n$-degenerate curve into $n$ double-degenerate curves and the $\mathbb{Z}_2$ topological phase would have the robustness by the same argument.

To confirm the above expectation, we consider the systems with PBC composed by three and four stacked chains, 
which are defined by
\begin{equation}
\mathcal{H}_{3\text{s}}(k)=
\begin{pmatrix}
\mathcal{H}(k) & \delta_0+i\boldsymbol{\delta}\cdot\boldsymbol{\sigma} & \\ 
\delta_0-i\boldsymbol{\delta}\cdot\boldsymbol{\sigma} & \mathcal{H}(k) &\delta_0+i\boldsymbol{\delta}\cdot\boldsymbol{\sigma}\\
 & \delta_0-i\boldsymbol{\delta}\cdot\boldsymbol{\sigma} & \mathcal{H}(k)
\end{pmatrix}
\end{equation}
and
\begin{equation}
\mathcal{H}_{4\text{s}}(k)=
\begin{pmatrix}
\mathcal{H}(k) & \delta_0+i\boldsymbol{\delta}\cdot\boldsymbol{\sigma} & & \\ 
\delta_0-i\boldsymbol{\delta}\cdot\boldsymbol{\sigma} & \mathcal{H}(k) &\delta_0+i\boldsymbol{\delta}\cdot\boldsymbol{\sigma} & \\
 & \delta_0-i\boldsymbol{\delta}\cdot\boldsymbol{\sigma} & \mathcal{H}(k) &\delta_0+i\boldsymbol{\delta}\cdot\boldsymbol{\sigma}\\
  & & \delta_0-i\boldsymbol{\delta}\cdot\boldsymbol{\sigma} & \mathcal{H}(k)
\end{pmatrix},
\end{equation}
respectively.
Note that both $\mathcal{H}_{3\text{s}}(k)$ and $\mathcal{H}_{4\text{s}}(k)$ exhibit spinful TRS$^{\dagger}$ symmetry with time reversal operators $\mathcal{T}_{\text{3s}}=I_3 \otimes \sigma_y$ and $\mathcal{T}_{\text{4s}}=I_4 \otimes \sigma_y$, respectively. Here, $I_n \ (n \in \mathbb{Z})$ represents an $n$-dimensional identity matrix. The definition of the $\mathbb{Z}_2$ invariant is the same as before [Eq. (\ref{eq:Z2 invariant})]. 

\begin{figure}
\centering
\includegraphics[width=\columnwidth]{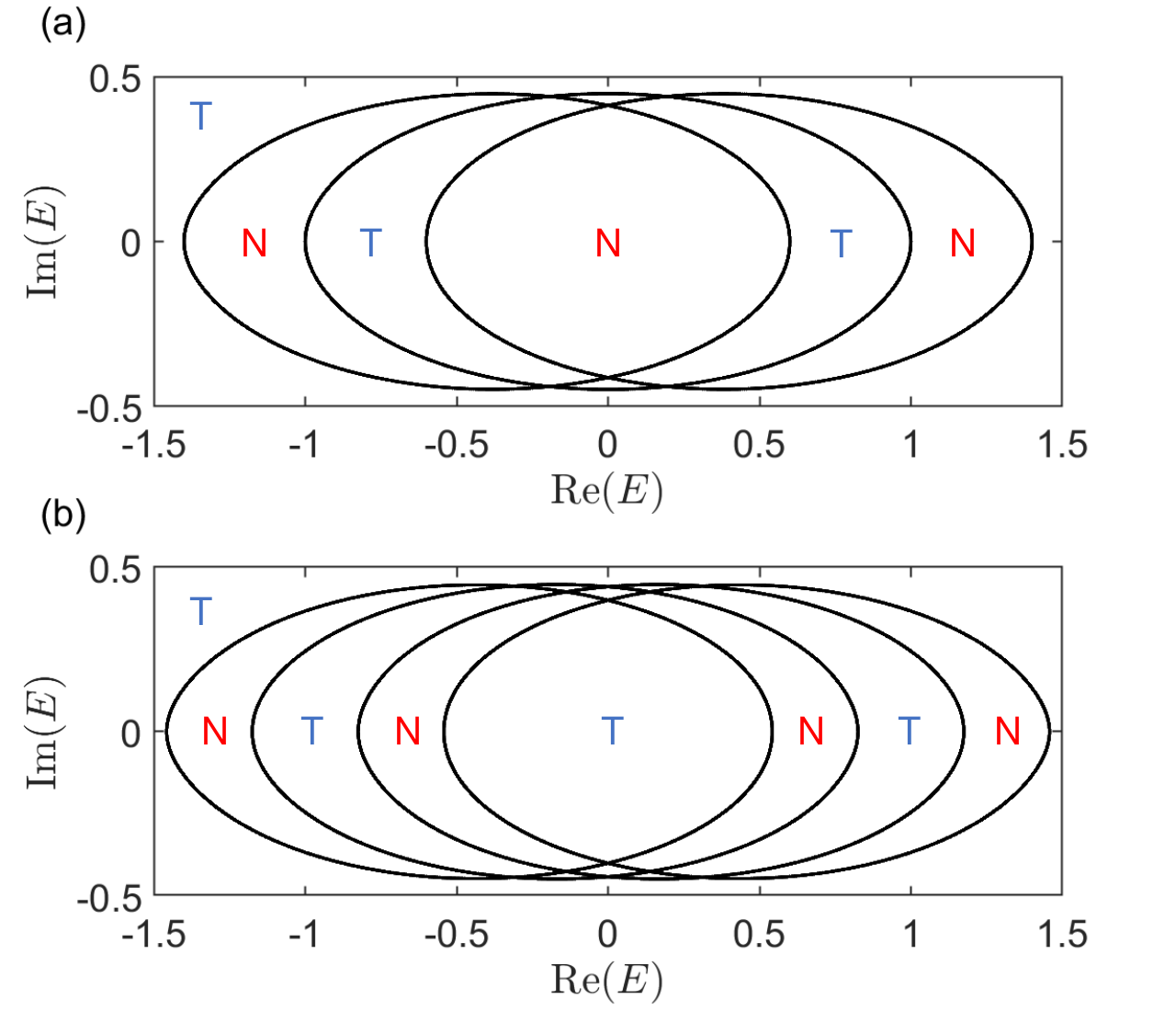}
\caption{Complex energy spectrum and $\mathbb{Z}_2$ index of (a) $\mathcal{H}_{3\text{s}}(k)$ and (b) $\mathcal{H}_{4\text{s}}(k)$ with parameters $t=0.5$, $\alpha_R=0.2$, $d_x=0.4$, $d_z=0.3$, $\delta_0=\delta_z=0.2$ and $\sigma_x=\sigma_y=0$. The capital T and N indicates that the $\mathbb{Z}_2$ topology is trivial or non-trivial, respectively, in the corresponding areas.}
\label{fig:multi-stacked system}
\end{figure}

The complex energy spectra of $\mathcal{H}_{3\text{s}}(k)$ and $\mathcal{H}_{4\text{s}}(k)$ are shown in Fig. \ref{fig:multi-stacked system}. We observe that PBC spectra of $\mathcal{H}_{3\text{s}}(k)$ and $\mathcal{H}_{4\text{s}}(k)$ split into 3 and 4 bands (ellipses) with double degeneracy, respectively. We also confirm that the $\mathbb{Z}_2$ invariant alternatively change the value when the point gap closes.
Moreover, for even-chain stacked systems, the energy at the origin is always topologically trivial, whereas for odd-chain stacked systems, the energy at the origin always shows non-trivia $\mathbb{Z}_2$ topology.

\subsection{Robust $\mathbb{Z}_2$ Topology: Hermitian System}
\label{sec:Hermitian system}

Finally, we show that the above mentioned robustness of $\mathbb{Z}_2$ topological phases in non-Hermitian systems against stacking is inherited to the Hermitian systems related by the Hermitization. 

Frist, we introduce a Hermitian Hamiltonian by using the stacked non-Hermitian Hamiltonian in Sec.\ \ref{sec:non-Hermitian system} through the Hermitization in Eq.\ (\ref{eq:Hermitization}) as
\begin{equation}
\begin{split}
H'_s(k)&
=\begin{pmatrix}
 0  &\mathcal{H}_s(k) - E_p \eta_0\otimes\sigma_0  \\
\mathcal{H}^\dagger_s(k) - E_p^* \eta_0\otimes\sigma_0  & 0
\label{eq:Hermitization2}
 \end{pmatrix}
\end{split}
\end{equation}
where $\mathcal{H}_{\text{s}}(k)$ is defined by Eq. (\ref{eq:stacked coupled HN}). We note that $E_p$ is a parameter standing for the reference point on the complex energy plain.
After a unitary transformation by $H_s(k)=U H^\prime_s(k) U^\dagger$ with $U=e^{i\frac{\pi}{4}\eta_0\otimes\tau_y\otimes\sigma_0}$, 
we obtain
\begin{equation}
\begin{split}
&H_s(k)=
\begin{pmatrix}
H_{11}(k) & H_{12} \\
H_{12}^\dagger & -H_{11}^T(-k)
\end{pmatrix},\label{eq:H_s(k)}
\\
&H_{11}(k)=
\begin{pmatrix}
h(k)-\text{Re}[E_p] & \delta_0\sigma_0 + i \boldsymbol{\delta} \cdot \boldsymbol{\sigma} \\
\delta_0\sigma_0 - i \boldsymbol{\delta} \cdot \boldsymbol{\sigma} & h(k)-\text{Re}[E_p] 
\end{pmatrix},\\
&H_{12}=
\begin{pmatrix}
i\boldsymbol{d}(k) \cdot \boldsymbol{\sigma} \sigma_y -i \text{Im}[E_p] & 0\\
0 & i\boldsymbol{d}(k) \cdot \boldsymbol{\sigma} \sigma_y -i \text{Im}[E_p] \\
\end{pmatrix},
\end{split}
\end{equation}
where $h(k)$ and $\boldsymbol{d}(k)$ are defined by Eq. \ref{eq:h(k)} and Eqs. \ref{eq:d(k)}, respectively. 
Here  we assume that $E_p$ is real. Rewriting $E_p$ as a parameter $\mu$ and rearranging the order of the basis,  Eq.\ (\ref{eq:H_s(k)}) can be written down  
\begin{equation}
\begin{split}
&H_s(k)=\\
&\begin{pmatrix}
h(k) -\mu &  i\boldsymbol{d}(k) \cdot \boldsymbol{\sigma} \sigma_y & \delta_0+i\boldsymbol{\delta}\cdot\boldsymbol{\sigma} & 0 \\
-i\boldsymbol{d}^*(k) \cdot \boldsymbol{\sigma} \sigma_y  & -h^{T}(-k) + \mu & 0 & -\delta_0-i\boldsymbol{\delta}\cdot\boldsymbol{\sigma} \\
\delta_0-i\boldsymbol{\delta}\cdot\boldsymbol{\sigma} & 0 & h(k) -\mu&  i\boldsymbol{d}(k) \cdot \boldsymbol{\sigma} \sigma_y\\
0 & -\delta_0+i\boldsymbol{\delta}\cdot\boldsymbol{\sigma} & -i\boldsymbol{d}^*(k) \cdot \boldsymbol{\sigma} \sigma_y  & -h^{T}(-k)+\mu
\end{pmatrix}.
\end{split}
\label{eq:H_s(k)2}
\end{equation}
The above Hamiltonian can be interpreted as the two stacked chains of the DIII Hermitonian Hamiltonian in Eq.\ (\ref{eq:rewritten Hamiltonian}), in which
each symmetry is given by
\begin{align}
\text{TRS}:\ &TH^*(k)T^{-1}=H(-k), \ T=\eta_0\otimes \tau_z\otimes \sigma_y,\\
\text{PHS}:\ &PH^*(k)P^{-1}=-H(-k), \ P=\eta_0\otimes \tau_y\otimes \sigma_y,\\
\text{CS}:\ &\Gamma H(k)\Gamma^{-1}=-H(k), \ \Gamma=\eta_0\otimes \tau_x\otimes \sigma_0.
\end{align}

We emphasize that we introduce $E_p$ as the reference point in Eq.\ (\ref{eq:Hermitization2}). In Sec.\ \ref{sec:model}, the reference point $E_p$ appears in Eq.\ (\ref{eq:Z2 invariant}) and we explain that if the non-Hermitian Hamiltonian is topologically non-trivial at the reference energy point $E_p$, the corresponding Hermitian Hamiltonian is also non-trivial in Fig.\ \ref{fig:spectrum of non-stacked system}. Then, in Sec.\ \ref{sec:non-Hermitian system}, we show that the $\mathbb{Z}_2$ point-gap topological phase remains robust against stacking if $E_p$ takes a proper values on the complex plain.  Combining these facts, we can predict that if $E_p$ locates at the non-trivial region for $\mathbb{Z}_2$ index of the non-Hermitian Hamiltonian $\mathcal{H}_\text{s}(k)$, the corresponding Hermitian DIII Hamiltonian $H_\text{s}(k)$ should also have non-trivial $\mathbb{Z}_2$ topological phase even in the stacked system. We verify this prediction below.

\begin{figure}
\centering
\includegraphics[width=\columnwidth]{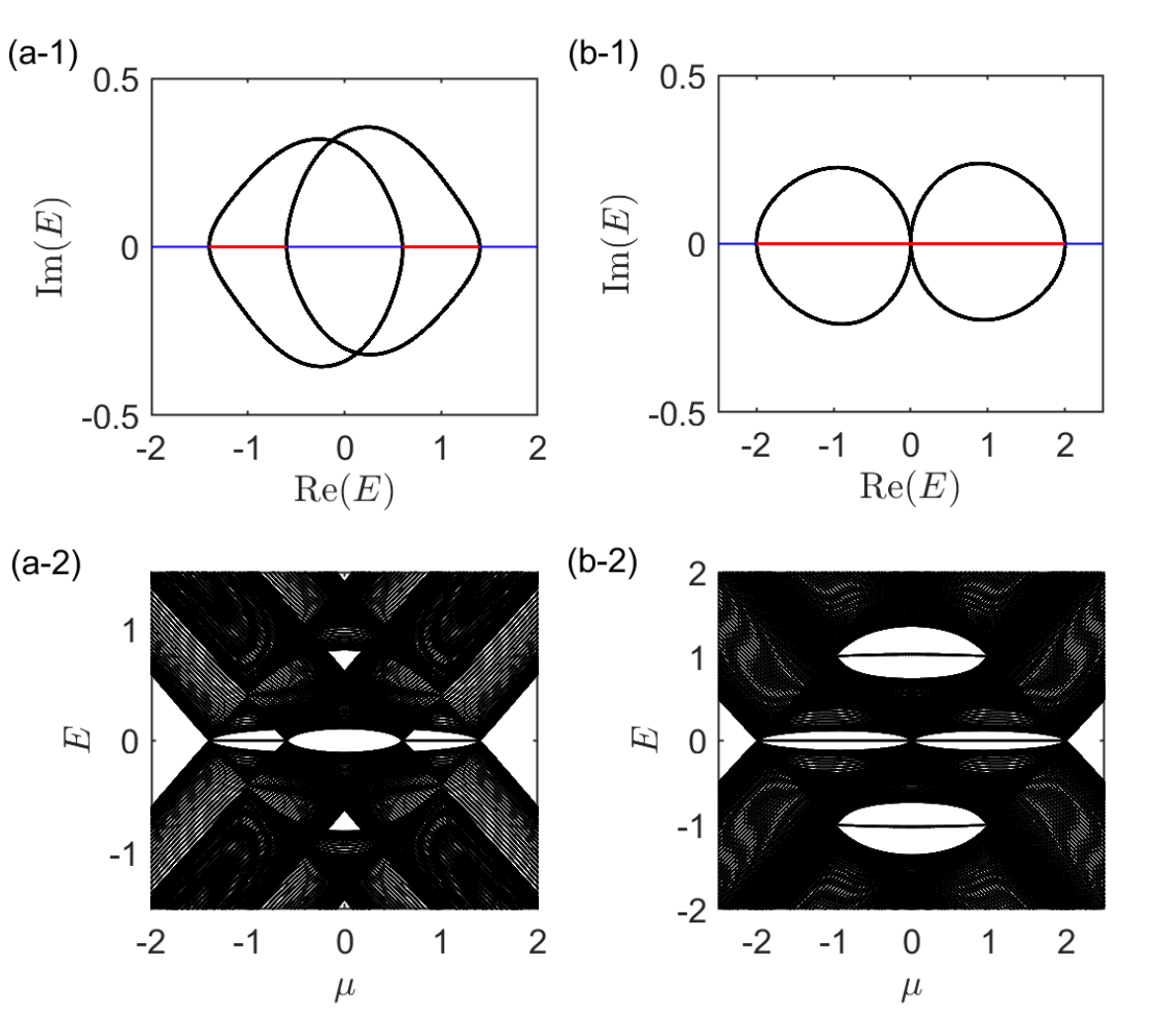}
\caption{Energy spectrum of (a-1,b-1) non-Hermitian stacked coupled HN model $\mathcal{H}_{\text{s}}(k)$ and (a-1,b-2) the corresponding stacked DIII superconductor $H_{\text{s}}(k)$, with $t=0.5$, $\alpha_R=0.2$, $d_x=0.4$, and $d_z=0.3$. Panels (a-1) and (a-2) correspond to the stacking strengths $\delta_i=0.2 \ (i=0,x,y,z)$, while panels (b-1) and (b-2) correspond to $\delta_i=0.5 \ (i=0,x,y,z)$. When $\mu$ takes values from red line segments in (a-1) or (b-1), the corresponding Hermitian system (a-2) or (b-2) exhibits zero-energy states.}
\label{fig:spectrum of stacked Hermitian system}
\end{figure}

While, in Sec.\ \ref{sec:non-Hermitian system}, $\delta_x=\delta_y=0$ is studied for simplicity, we consider all the coupling terms here.
Figures \ref{fig:spectrum of stacked Hermitian system} (a-1) and (b-1) shows the PBC spectrum of the stacked non-Hermitian Hamiltonian $\mathcal{H}_{\text{s}}(k)$ with $\delta_{i}=0.2$ and $0.5$ for $i=0,x,y,z$, respectively.
Similar to Fig. \ref{fig:spectrum of stacked coupled HN}, we observe two closed curves intersecting each other at $\delta_{i}=0.2$ and attaching at $E=0$ at $\delta_{i}=0.5$, 
which arise from level repulsion. 
Note that the closed curves are not ellipses because of contributes of $\delta_x$ and $\delta_y$.
Calculating the $\mathbb{Z}_2$ invariant of $\mathcal{H}_{\text{s}}(k)$, we confirm that only the non-overlapping region exhibits non-trivial $\mathbb{Z}_2$ topology. 

Now we study the corresponding Hermitian DIII Hamiltonian in Eq.\ (\ref{eq:H_s(k)2}) with OBC and verify the stability of the zero-energy states originating the from $\mathbb{Z}_2$ topological phase. 
To begin with, we note that the energy-gap around $E~0$ of the Hamiltonian close at $\displaystyle |\mu| = 2t \pm \sqrt{\sum_{i=0,x,y,z} \delta_i^2}$ which can be obtained as the fact that the Hamiltonian closes the gap at $k=0,\pi$.
The OBC spectra are shown in Figs. \ref{fig:spectrum of stacked Hermitian system} (a-2) and (b-2).
As depicted in Fig.\ref{fig:spectrum of stacked Hermitian system} (a-2) for $\delta_i=0.2$, for $|\mu| < 2t-\sqrt{\sum_{i}\delta_i^2}=0.9$, $H_s(k)$ exhibits trivial $\mathbb{Z}_2$ topology, and  no states appear in the energy gap around $E=0$. On the other had, for $2t-\sqrt{\sum_i\delta_i^2} < |\mu| < 2t+\sqrt{\sum_i\delta_i^2}$, $H_s(k)$ exhibits non-trivial $\mathbb{Z}_2$ topology, and fourfold-degenerate zero-energy edge states emerge even in the stacked DIII system under OBC. This demonstrates the robustness of $\mathbb{Z}_2$ topology and zero-energy edge states in a stacked system. Particularly, when the stacking strengths reach a critical value such that $\sum_{i}\delta_i^2\ =4t^2$, all energy points inside the PBC spectra exhibit non-trivial $\mathbb{Z}_2$ topology as shown in Fig.\ \ref{fig:spectrum of stacked Hermitian system} (b-1). Accordingly, for all $\mu$ satisfying $0<|\mu|<4t$ ($\mu\neq0$), the DIII Hermitian Hamiltonian exhibits zero-energy states as shown in Fig. \ref
 {fig:spectrum of stacked Hermitian system} (b-2). 

We can naively extend the above result to an $n$-chain stacked Hermitian system.
Taking into account the results in Fig.\ \ref{fig:AppA1} for the multi-stacked non-Hermitian, trivial and non-trivial $\mathbb{Z}_2$ invariant of the corresponding Hermitian Hamiltonian alternatively appears as the function of the (real) chemical potential $\mu$ which moves on the real axis in in Fig.\ \ref{fig:AppA1}.

\subsection{Revisiting previous work}

So far, we explain that $\mathbb{Z}_2$ point-gap topology is robust against stacking due to level repulsion and then the corresponding Hamiltonian derived through the Hermitization inherits the robustness of $\mathbb{Z}_2$ topology. We demonstrated the robustness by focusing on the specific Hamiltonian belonging to class DIII.

As we mentioned in the introduction, several works reported the non-trivial $\mathbb{Z}_2$ topology in the stacked Hermitian systems\cite{PhysRevB.84.054532,PhysRevB.92.235407,PhysRevB.94.235414}, while the reason has not yet been understood.
Here we clarify that these results can by explained by non-Hermitian level repulsion. 

Here we focus on Refs. \cite{PhysRevB.92.235407,PhysRevB.94.235414} in which 
the stacking of the 2D QSH insulator approaching to the 3D Wilson-Dirac-type Hamiltonian is studied. It is shown that the $\mathbb{Z}_2$ invariant defined for the 2D QSH insulator remains robust against the number of stacked systems as a function of a mass parameter (equivalent to $\mu$ in our model).
While the QSH insulator generally belongs to class AII, the on-site random potential term is ignored for the calculation of the $\mathbb{Z}_2$ invariant. 
In this case, the Hamiltonian recovers chiral symmetry and belongs to class DIII which makes it possible to derive the corresponding non-Hermitian Hamiltonian belonging to class AII$^\dagger$.
Thereby, the robustness of $\mathbb{Z}_2$ invariant in Refs. \cite{PhysRevB.92.235407,PhysRevB.94.235414} can be explained by level repulsion of the non-Hermitian Hamiltonian.

\section{Summary and Discussion}
\label{sec:summary}

As a summary, in this work, we proposed the perspective that non-Hermitian level repulsion is the fundamental reason behind the robustness of $\mathbb{Z}_2$ topology in Hermitian systems. Our main methodology involves connecting the topological properties of Hermitian systems with those of non-Hermitian systems through a transformation called Hermitization. 
While $\mathbb{Z}_2$ invariant of stacked Hermitian systems becomes trivial or non-trivial depending on system parameters, the energy region of non-trivial $\mathbb{Z}_2$ invariant in non-Hermitian Hamiltonian originates from level repulsion which can be easily understood since only the two-fold degeneracy is guaranteed by Kramers pairs.


To demonstrate our general argument, we studied the 1D DIII topological superconductor with chiral symmetry. Through unitary transformations and inverse Hermitization, we establish a connection with a stacked coupled Hatano-Nelson model in the AII$^{\dagger}$ symmetry class. As predicted, the energy spectrum of the stacked non-Hermitian system shows level repulsion and we confirmed the existence of $\mathbb{Z}_2$ non-trivial regions on the energy complex plain. Then, we clarified that the corresponding stacked Hermitian system also become $\mathbb{Z}_2$ topologically non-trivial when the chemical potential $\mu$ locates at the real axis on the non-trivial regions. 

Furthermore, we showed that the robustness of $\mathbb{Z}_2$ topological invariant of the stacked QSH insulators studied in Ref.\ \cite{PhysRevB.92.235407,PhysRevB.94.235414} can be well explained by level repulsion of the corresponding non-Hermitian system since the system without on-site disordered potential has chiral symmetry and belongs to class DIII in 2D. We emphasize again that the non-trivial $\mathbb{Z}_2$ topology can be easily understood as the consequence of level repulsion in case of non-Hermitian Hamiltonians.

\chdeleted{y, the model studied in Ref. [67] has chiral symmetry since the effect of disorder is ignored. Both of these works show consistent results to ours while different symmetry classes and dimension are considered. Therefore, We believe our approach is universal.}

At last, we note the several open questions.
As we explained that Ref.\ \cite{PhysRevB.84.054532} which study the stacked Kitaev chains belonging to class D also observed the robustness of $\mathbb{Z}_2$ topological invariant against disorder, while there is no chiral symmetry in class D. 
While the \chreplaced{K}{k}itaev model can be transformed into class BDI which possesses chiral symmetry after a proper gauge transformation, the corresponding non-Hermitian system belongs to class AI and has $\mathbb{Z}$ point-gap topology. Thereby, we cannot apply our theory to this system. The system studied by Ref.\ \cite{PhysRevB.92.235407,PhysRevB.94.235414} also lose chiral symmetry if the system has on-site disorder potential, nevertheless the conductance is well quantized. These results might be able to explained by generalizing the current theory which will be addressed in the future work.

\section*{Acknowledgements}

We thank Y. Asano, K.-I. Imura, T. Sato, A. Sasaki, and K. Yakubo for their helpful discussions. Especially, we appreciate K.-I. Imura for introducing us to the result in Ref.\ \cite{PhysRevB.94.235414}. Z.J. was supported by JST SPRING (Grant No. JPMJSP2119). This work was supported by JSPS KAKENHI (Grants No. JP20H01828, No. JP22K03463, No. JP23K22411, No. JP24K00545,	No. JP24K00569). M.S. was supported by JST CREST Grant No. JPMJCR19T2. H.O. was supported by JST PRESTO Grant No. JPMJPR2454.

\appendix

\section{Stacking of A Common Coupled Hatano-Nelson Model}
\label{sec:appendixA}

In Sec. \ref{sec:non-Hermitian system}, we demonstrate the robustness of $\mathbb{Z}_2$ invariant due to level repulsion by using the non-Hermitian Hamiltonian derived from the Hermitian Hamiltonian of the DIII superconductor through the inverse Hermitization. In this Appendix, we show that the results are general properties by using the other model which is originally introduced in \cite{PhysRevLett.124.086801}. The Hamiltonian is given by 
\begin{equation}
\mathcal{H}_{\text{c}}(k)=\begin{pmatrix} \mathcal{H}_{\text{HN}}(k) & 2\Delta\sin k \\ 2\Delta\sin k & \mathcal{H}_{\text{HN}}^{T}(-k) \end{pmatrix},
\end{equation}
where
\begin{equation}
\mathcal{H}_{\text{HN}}(k)=(t+g)e^{ik}+(t-g)e^{-ik}
\end{equation}
describes the Hatano-Nelson model without disorder. The parameters $t$ and $g$ represent the asymmetric hopping terms, while $\Delta$ denotes the coupling strength. 
Such a system belongs to the AII$^{\dagger}$ symmetry class, and $\mathbb{Z}_2$ point gaps remain open as long as $g \neq 0$.

Similar to Eq. (\ref{eq:stacked Hamiltonian}), we define the stacked system $\mathcal{H}_{\text{s}}(k)$ as
\begin{equation}
\mathcal{H}_{\text{s}}(k)=\begin{pmatrix} \mathcal{H}_{\text{c}}(k) & \delta_0 \sigma_0+i\boldsymbol{\delta} \cdot \boldsymbol{\sigma} \\ \delta_0 \sigma_0-i\boldsymbol{\delta} \cdot \boldsymbol{\sigma} & \mathcal{H}_{\text{c}}(k) \end{pmatrix},
\end{equation}
where the parameters are defined as in Eq. (\ref{eq:stacked coupled HN}). For simplicity, we still consider the case where $\delta_x = \delta_y = 0$, and the energy spectrum is solved as follows:
\begin{equation}
E_{\eta\epsilon}(k)=2t\cos k+i\eta r(k)e^{i\varepsilon\phi(k)},
\label{eq:eigenvalue of stacked coupled HN model appendix}
\end{equation}
where
\begin{equation}
\begin{split}
r(k)=&\left\{ \left[ 4(g^2-\Delta^2)\sin^2 k-(\delta_0^2+\delta_z^2) \right]^2 \right. \\
& \left. -(4\sin k)^2\left[ \delta_0^2(g^2-\Delta^2)+\delta_z^2g^2 \right] \right\}^{\frac{1}{4}},
\end{split}
\end{equation}
\begin{equation}
\phi(k)=\frac{1}{2}\arctan \frac{4\sin k\sqrt{\delta_0^2(g^2-\Delta^2)+\delta_z^2g^2}}{(g^2-\Delta^2)\sin^2 k-(\delta_0^2+\delta_z^2)}\in\left( -\frac{\pi}{2},\frac{\pi}{2} \right).
\end{equation}
Here, $\eta,\varepsilon=\pm$ represent the four energy bands $E_{\pm\pm}$ of $\mathcal{H}_{\text{s}}(k)$.
We notice that for $\delta_0\neq 0$ or $\delta_z\neq 0$, the complex energy spectrum of $\mathcal{H}_{\text{c}}(k)$ splits into a double degeneracy because of $E_{++}\neq E_{-+}$.

\begin{figure}
\centering
\includegraphics[width=\columnwidth]{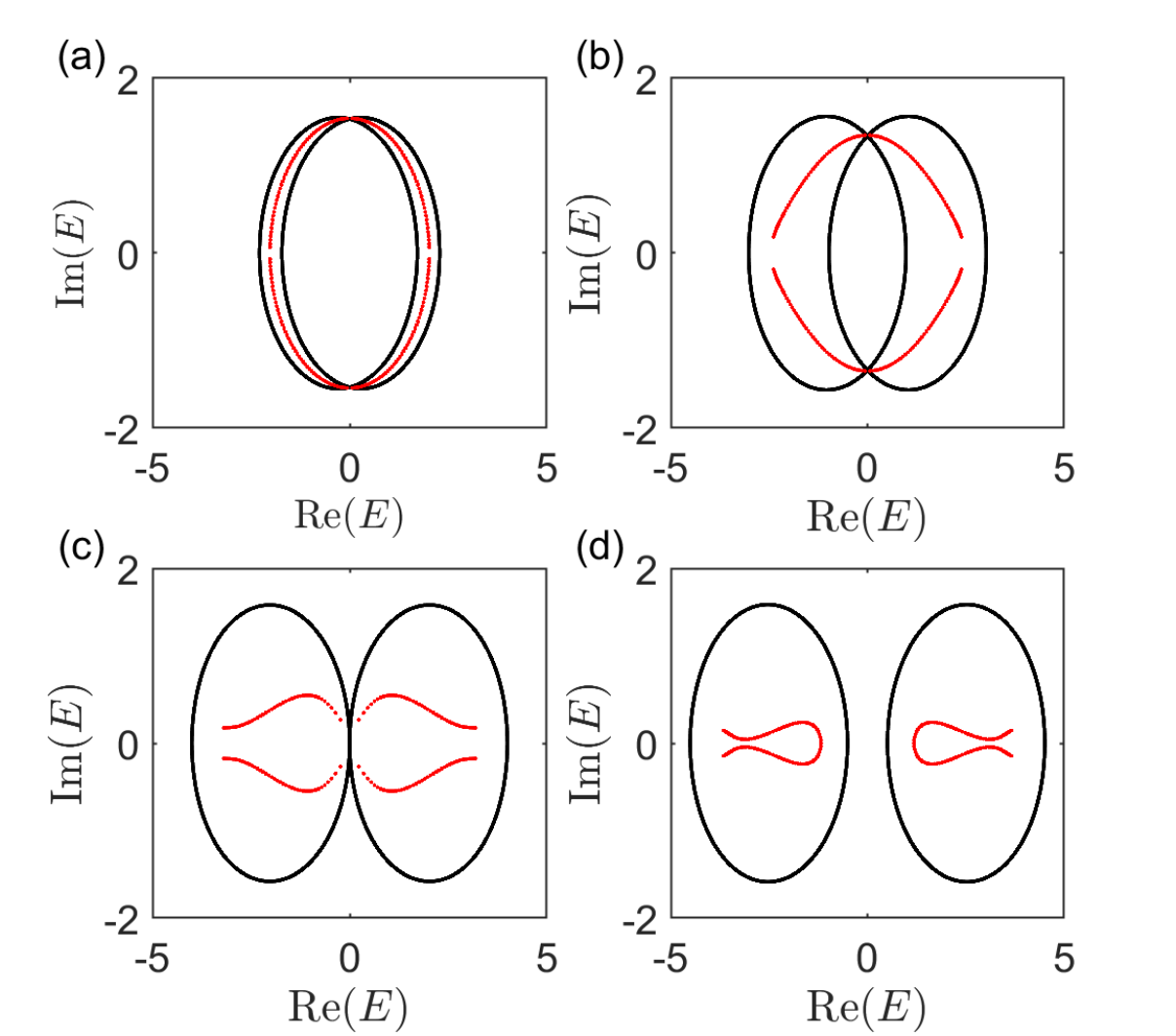}
\caption{ Complex energy sppectrum of stacked coupled HN model $\mathcal{H}_{\text{s}}$ under PBC (black curves) and OBC (rea curves), with parameters $t=1$, $g=0.8$, $\Delta=0.2$ and $\delta_0=0.2$ (a) $\delta_z=0.2$ (b) $\delta_z=1$ (c) $\delta_z=\sqrt{3.96}$ (d) $\delta_z=2.5$.}
\label{fig:AppA1}
\end{figure}

\begin{figure}
\centering
\includegraphics[width=\columnwidth]{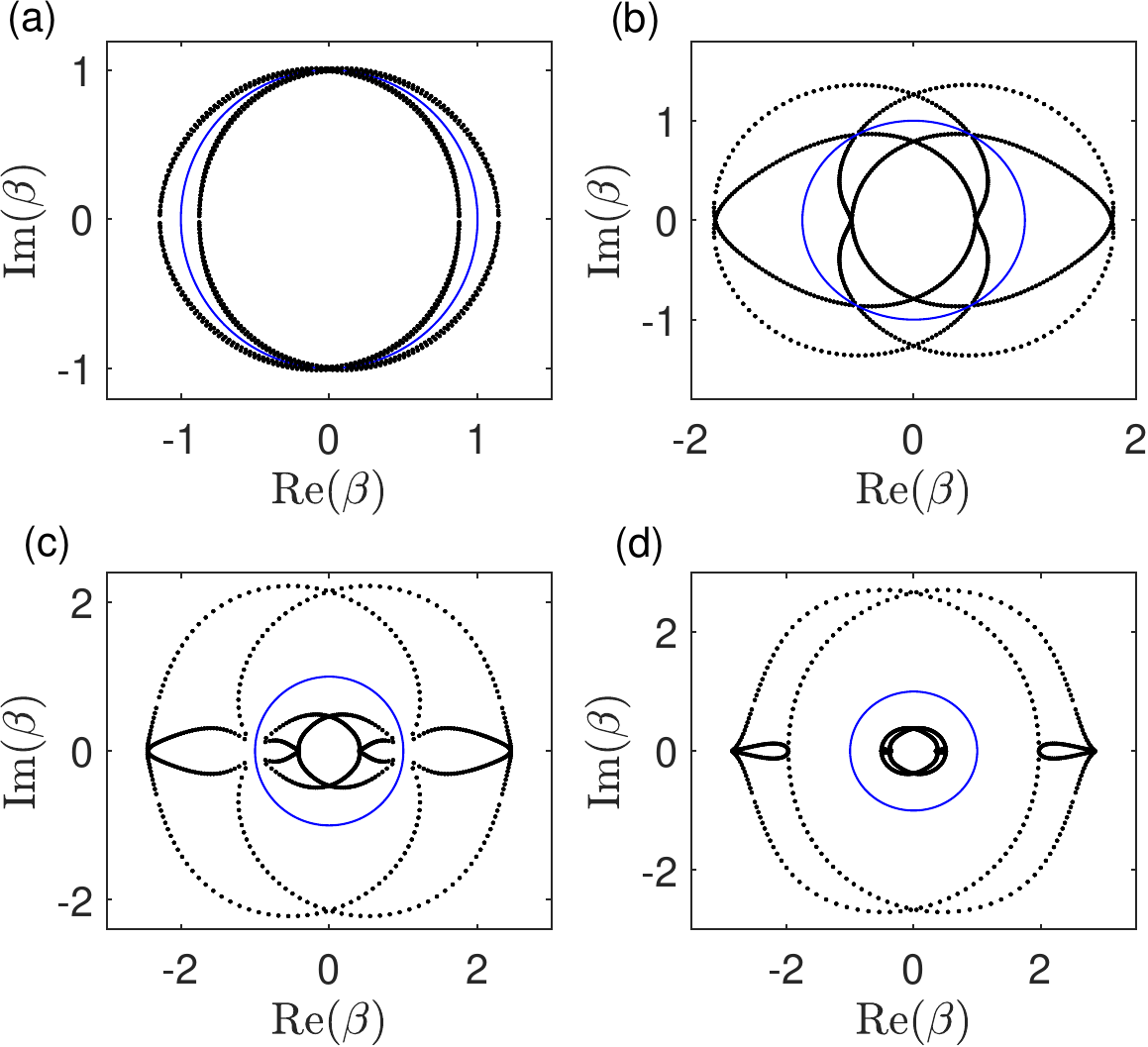}
\caption{ GBZ (black dots) and BZ (blue unit circle) of stacked coupled HN model $\mathcal{H}_{\text{s}}$, with parameters $t=1$, $g=0.8$, $\Delta=0.2$ and $\delta_0=0.2$ (a) $\delta_z=0.2$ (b) $\delta_z=1$ (c) $\delta_z=\sqrt{3.96}$ (d) $\delta_z=2.5$.}
\label{fig:AppA2}
\end{figure}

Figure \ref{fig:AppA1} shows the PBC and OBC spectra of the non-Hermitian Hamiltonian $\mathcal{H}_{\text{s}}(k)$. The OBC spectrum is derived from the GBZ, as shown in Fig. \ref{fig:AppA2} due to strong finite-size effects. From Fig. \ref{fig:AppA1}, we observe that the PBC spectrum forms closed curves, while the OBC spectrum forms arcs with no enclosed area, always lying within the $\mathbb{Z}_2$ non-trivial region of the PBC spectrum. This indicates the robustness of the NHSE in this model against stacking, similar to the model discussed in Sec. \ref{sec:non-Hermitian system}.

\section{Influence of Stacking Configuration on Topology}
\label{app:app2}

In this Appendix, we explain why we need to consider, at least, two terms of the stacking terms $\delta_0 + i\boldsymbol{\delta}\cdot\boldsymbol{\sigma}$ when defining the stacked system in Sec. \ref{sec:Hermitian system}. We show that the Hamiltonian is decoupled into two independent ones and the symmetry class may also change if we consider only a single stacking term. 

\subsection{$\delta_0\neq0$, $\delta_x=\delta_y=\delta_z=0$}

First, we consider the simplest case by focusing only on a spin-independent stacking term, $\delta_0$. The stacked non-Hamiltonian, $\mathcal{H}_{\text{s}0}(k)$, is expressed as follows:
\begin{equation}
\mathcal{H}_{\text{s}0}(k)=\begin{pmatrix} \mathcal{H}(k) & \delta_0\sigma_0 \\ \delta_0\sigma_0 & \mathcal{H}(k)\end{pmatrix}.
\end{equation}
By applying a unitary transformation with $\mathcal{U}_0 = e^{-i\frac{\pi}{4}\eta_y \otimes \sigma_0}$, we can obtain a diagonal block Hamiltonian
\begin{equation}
\begin{split}
\mathcal{H}'_{\text{s}0}(k)&=\mathcal{U}_0\mathcal{H}_{\text{s}0}(k)\mathcal{U}_0^{-1}\\
&=\begin{pmatrix} \mathcal{H}(k)-\delta_0\sigma_0 &  \\  & \mathcal{H}(k)+\delta_0\sigma_0\end{pmatrix}\\
&=\begin{pmatrix} \mathcal{H}_{0}^{(1)}(k) & \\ & \mathcal{H}_{0}^{(2)}(k) \end{pmatrix}.
\end{split}
\end{equation}

\begin{figure}
\centering
\includegraphics[width=\columnwidth]{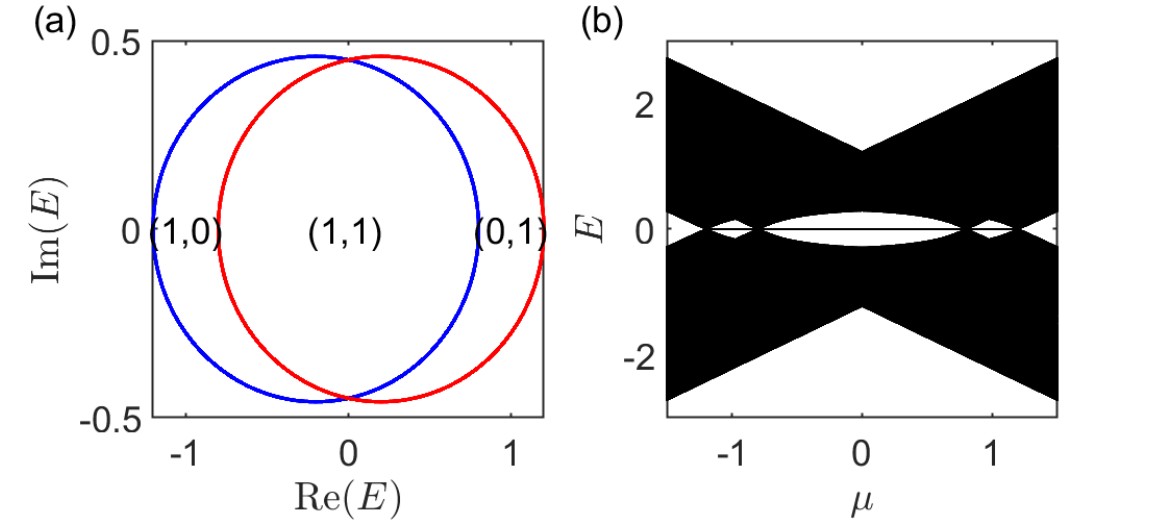}
\caption{(a) Energy spectrum of $\mathcal{H}_{0}^{(1)}(k)$ (blue curve) and $\mathcal{H}_{0}^{(2)}(k)$ (red curve). The numbers present the $\mathbb{Z}_2$ topological number of $\mathcal{H}_{0}^{(1)}(k)$ and $\mathcal{H}_{0}^{(2)}(k)$, respectively. (b) Real energy spectrum of the corresponding stacked Hermitian system, with $t=0.5$, $\alpha_R=0.2$, $d_x=0.4$, $d_z=0.3$, $\delta_0=0.2$ and $\delta_x=\delta_y=\delta_z=0$. The system exhibits zero-energy edge states in both regions $|\mu| < 0.8$ and $0.8 < |\mu| < 1.2$.}
\label{figure:spectrum0}
\end{figure}

$\mathcal{H}_{0}^{(1)}(k)$ and $\mathcal{H}_{0}^{(2)}(k)$ represent two decoupled systems belonging to AII$^{\dagger}$ class, each acquiring non-trivial $\mathbb{Z}_2$ topology at the energy points enclosed by their own spectrum [Fig.\ref{figure:spectrum0}(a)]. As illustrated in the Fig.\ref{figure:spectrum0}(b), the corresponding Hermitian Hamiltonian $H_{s0}$ exhibits zero-energy edge states in both regions $|\mu| < 2t - |\delta_0|$ and $2t - |\delta_0| < |\mu| < 2t + |\delta_0|$, protected by non-trivial $\mathbb{Z}_2$ topology.

\subsection{$\delta_z\neq0$, $\delta_0=\delta_x=\delta_y=0$}

Here, we consider the case of focusing solely on the spin-dependent stacking term $\sigma_z$. The stacked non-Hermitian Hamiltonian, $\mathcal{H}_{\text{s}z}(k)$, is expressed as follows:
\begin{equation}
\mathcal{H}_{\text{s}z}(k)=\begin{pmatrix} \mathcal{H}(k) & i\delta_z\sigma_z \\  -i\delta_z\sigma_z  & \mathcal{H}(k)\end{pmatrix},
\end{equation}
By applying a unitary transformation with $\mathcal{U}_z = e^{-i\frac{\pi}{4} \eta_x \otimes \sigma_0}$, we can obtain a diagonal block Hamiltonian
\begin{equation}
\begin{split}
\mathcal{H}'_{\text{s}z}(k)&=\mathcal{U}_z\mathcal{H}_{\text{s}z}(k)\mathcal{U}_z^{-1}\\
&=\begin{pmatrix} \mathcal{H}(k)-\delta_z  \sigma_z &  \\  & \mathcal{H}(k)+\delta_z\sigma_z\end{pmatrix}\\
&=\begin{pmatrix} \mathcal{H}_{z}^{(1)}(k) & \\ & \mathcal{H}_{z}^{(2)}(k) \end{pmatrix}.
\end{split}
\end{equation}

$\mathcal{H}_{z}^{(1)}(k)$ and $\mathcal{H}_{z}^{(2)}(k)$ are two decoupled systems, both belonging to the class A symmetry class, without TRS, PHS, or CS. Such systems achieve $\mathbb{Z}$ point-gap topology in 1D systems.

\begin{figure}
\centering
\includegraphics[width=\columnwidth]{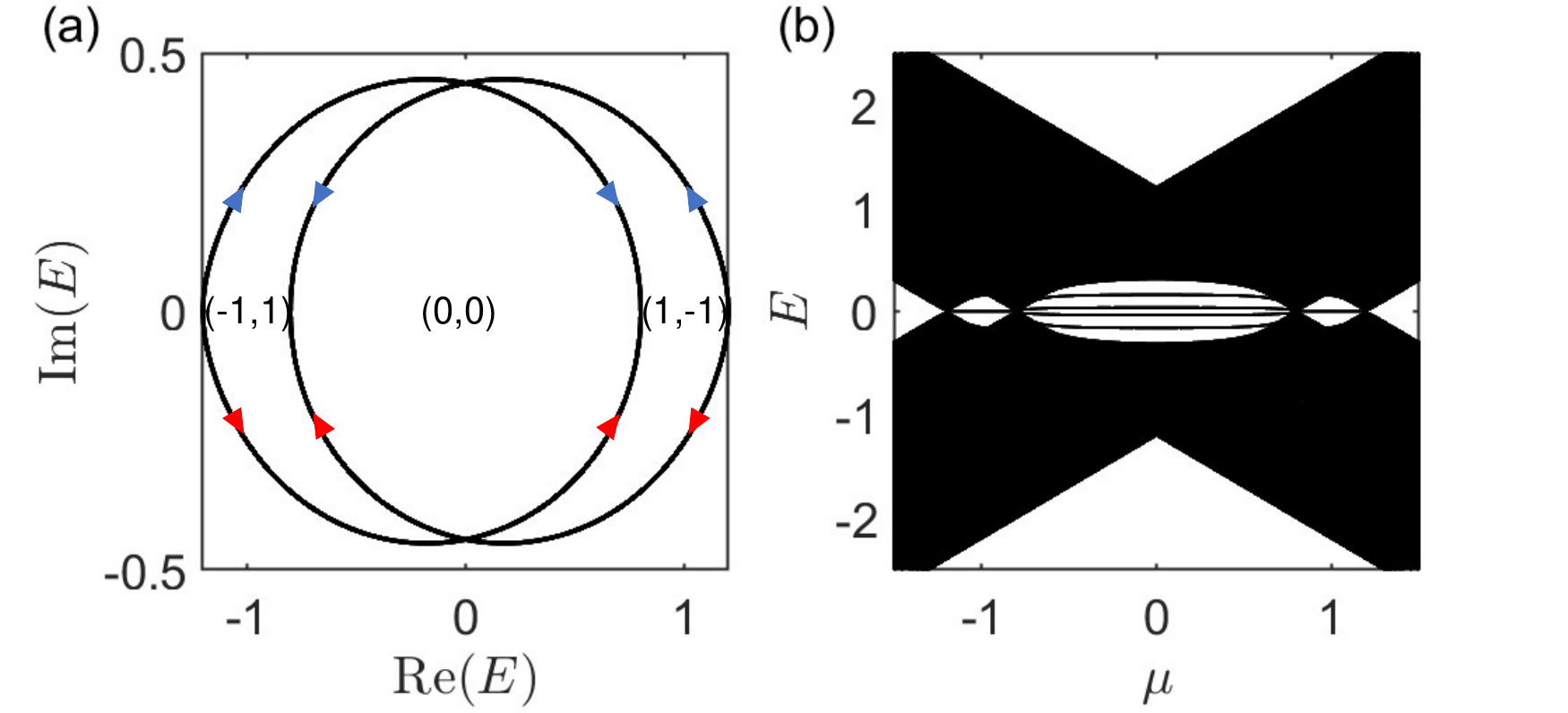}
\caption{Energy spectrum of (a) coincident $\mathcal{H}_{z}^{(1)}(k)$ and $\mathcal{H}_{z}^{(2)}(k)$ and (b) the corresponding Hermitian system, with $t=0.5$, $\alpha_R=0.2$, $d_x=0.4$, $d_z=0.3$, $\delta_z=0.2$ and $\delta_0=\delta_x=\delta_y=0$. The numbers in (a) indicate the winding number of $\mathcal{H}_{z}^{(1)}(k)$ and $\mathcal{H}_{z}^{(2)}(k)$ in the corresponding areas, while the direction of rotation of the energy spectrum is marked by blue arrows ($\mathcal{H}_{z}^{(1)}(k)$) and red arrows ($\mathcal{H}_{z}^{(2)}(k)$). The Hermitian system $H_{\text{s}z}$ exhibits zero-energy edge states in  the region $0.8 < |\mu| < 1.2$.}
\label{figure:spectrumz}
\end{figure}

As shown in Fig. \ref{figure:spectrumz}, the energy spectra of $\mathcal{H}_{z}^{(1)}(k)$ and $\mathcal{H}_{z}^{(2)}(k)$ coincide with each other, given by $E_{\eta,\varepsilon}=2t\cos{k}+\eta\sqrt{(id_x\sin{k}+\varepsilon d_z)^2+(\alpha_R^2-d_z^2)\sin{k}^2}$, where $\eta,\varepsilon=\pm 1$. However, they consistently exhibit opposite winding numbers at the same energy point, leading to the emergence of zero-energy edge states in the region $2t - |\delta_z| < |\mu| < 2t + |\delta_z|$, protected by non-trivial $\mathbb{Z}$ topology, but not $\mathbb{Z}_2$ topology.

\subsection{$\delta_x\neq0$, $\delta_0=\delta_y=\delta_z=0$}
\label{subsection:deltax}
In the case of only the spin-dependent stacking term $\delta_x \neq 0$, we employ the same method as when $\delta_z \neq 0$. The stacked Hamiltonian, $\mathcal{H}_{\text{s}x}(k)$, is expressed as follows:
\begin{equation}
\mathcal{H}_{\text{s}x}(k)=\begin{pmatrix} \mathcal{H}(k) & i\delta_x\sigma_x \\  -i\delta_x\sigma_x  & \mathcal{H}(k)\end{pmatrix},
\end{equation}
by a unitary transform by $\mathcal{U}_x=e^{-i\frac{\pi}{4}\eta_x\otimes\sigma_0}$, we can get a diaginal block Hamiltonian
\begin{equation}
\begin{split}
\mathcal{H}'_{\text{s}x}(k)&=\mathcal{U}_x\tilde{H}_{\text{s}x}(k)\mathcal{U}_x^{-1} \\
&=\begin{pmatrix} \mathcal{H}(k)-\delta_x  \sigma_x &  \\  & \mathcal{H}(k)+\delta_x\sigma_x\end{pmatrix}\\
&=\begin{pmatrix} \mathcal{H}_{x}^{(1)}(k) & \\ & \mathcal{H}_{x}^{(2)}(k) \end{pmatrix}.
\end{split}
\end{equation}
Here, neither $\mathcal{H}_{x}^{(1)}(k)$ nor $\mathcal{H}_{x}^{(2)}(k)$ possess any symmetries, which results in A symmetry class. Since the symmetry of $\mathcal{H}_{\text{s}x}$ is the same as that of $\mathcal{H}_{\text{s}z}$, $H_{\text{s}x}$ also achieves zero-energy edge states in the region $2t - |\delta_x| < |\mu| < 2t + |\delta_x|$, protected by non-trivial $\mathbb{Z}$ topology.

\subsection{$\delta_y\neq0$, $\delta_0=\delta_x=\delta_z=0$}

\begin{figure}
\centering
\includegraphics[width=\columnwidth]{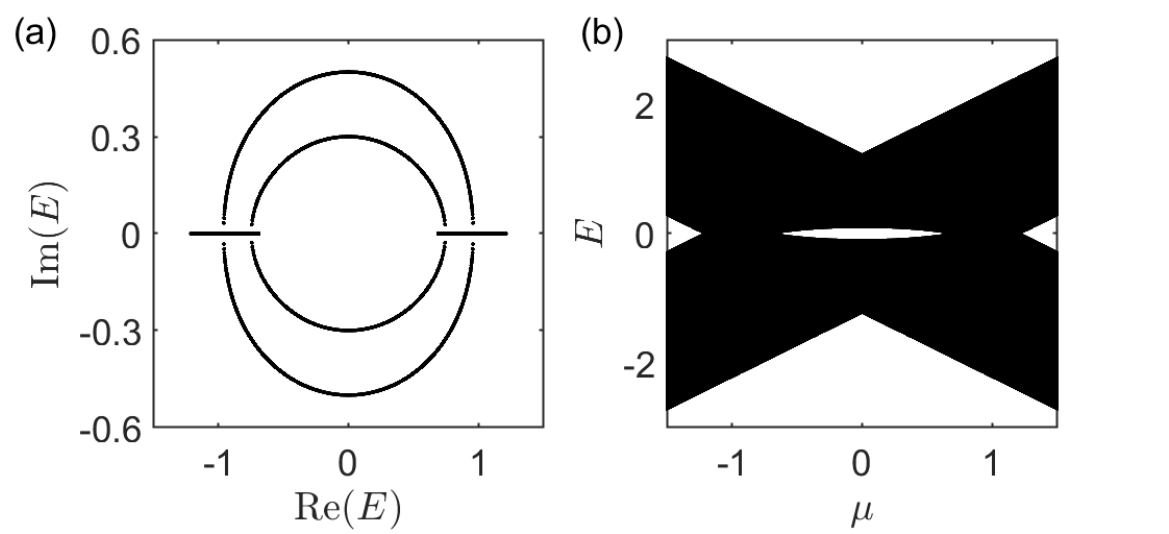}
\caption{Energy spectrum of (a) non-Hermitian Hamiltonian $\mathcal{H}_{\text{s}y}(k)$ and (b) stacked Hermitian Hamiltonian $H_{\text{s}y}$, with $t=0.5$, $\alpha_R=0.2$, $d_x=0.4$, $d_z=0.3$, $\delta_y=0.2$ and $\delta_0=\delta_x=\delta_z=0$.  The point-gap  is closed in the region of $0.8 < |\mu| < 1.2$, and no zero-energy edge states emerge in Hermitian system $H_{\text{s}y}$.}
\label{figure:spectrumy}
\end{figure}

Here, we focus on the spin-dependent stacking term, $\sigma_y$. The corresponding stacked non-Hamiltonian, $\mathcal{H}_{\text{s}y}(k)$, is expressed as follows:
\begin{equation}
\mathcal{H}_{\text{s}y}(k)=\begin{pmatrix} \mathcal{H}(k) & i\delta_y\sigma_y \\ -i\delta_y\sigma_y & \mathcal{H}(k)\end{pmatrix},
\end{equation}
by a unitary transform by $\mathcal{U}_y=e^{-i\frac{\pi}{4}\eta_x\otimes\sigma_0}$, we can get a diaginal block Hamiltonian
\begin{equation}
\begin{split}
\mathcal{H}'_{sy}(k)&=\mathcal{U}_y\mathcal{H}_{\text{s}y}(k)\mathcal{U}_y^{-1}\\
&=\begin{pmatrix} \mathcal{H}(k)-\delta_y\sigma_y &  \\  & \mathcal{H}(k)+\delta_y\sigma_y\end{pmatrix}\\
&=\begin{pmatrix} \mathcal{H}_{y}^{(1)}(k) & \\ & \mathcal{H}_{y}^{(2)}(k) \end{pmatrix}.
\end{split}
\end{equation}
Similar to the case discussed in Appendix \ref{subsection:deltax}, neither $\mathcal{H}_{y}^{(1)}(k)$ nor $\mathcal{H}_{y}^{(2)}(k)$ possess any symmetries. The eigenenergies can be obtained by solving the characteristic equation, as shown below:
\begin{equation}
E_{\eta,\varepsilon}=2t\cos{k}+\eta\sqrt{(\alpha_R\sin{k}+\varepsilon \delta_y)^2-(d_x^2+d_y^2)\sin^2{k}},
\end{equation}
where $\eta,\epsilon=\pm1$.
There must be a critical value $k_c$, such that for $k\in(0,k_c)$, $(\alpha_R\sin{k}\pm \delta_y)^2-(d_x^2+d_y^2)\sin^2{k}>0$, leading to the emergence of real eigenerengies $E\in(\pm 2t-|\sigma_y|,\pm 2t+|\sigma_y|)$. As shown in Fig. \ref{figure:spectrumy}(a), the energy gap always closes for $2t-|\delta_y|<|\mu|<2t+|\delta_y|$ in our model. As for the region $|\mu|<2t-|\delta_y|$, there is no zero-energy edge states due to the trivial topology [see Fig. \ref{figure:spectrumy}(b)].

In summary of Appendix \ref{app:app2}, if we introduce only the spin-independent stacking term $\delta_0$, the non-Hermitian Hamiltonian can be represented by two decoupled subsystems with $\mathbb{Z}_2$ point-gap topology using the unitary operator $e^{-i\frac{\pi}{4}\eta_y \otimes \sigma_0}$. If we introduce only $\sigma_{x}$ or $\sigma_z$, the non-Hermitian Hamiltonian can be achieved by two decoupled subsystems with opposite winding numbers using the unitary operator $e^{-i\frac{\pi}{4}\eta_x \otimes \sigma_0}$, and the edge states are protected by non-trivial $\mathbb{Z}$ topology. If we introduce only $\sigma_y$ stacking, the point gap of non-Hermitian Hamiltonian will close on the real axis,  causing the zero-energy states to disappear. By introducing both the stacking term $\delta_0$ and $\delta_{x/y/z}$ simultaneously, we can avoid the system decoupling into two subsystems and ensure that the zero-energy edge states are protected by $\mathbb{Z}_2$ topology.

\bibliographystyle{jiang-apsrev4-2}
\bibliography{template_20240921}

\end{document}